\documentclass{article}
\usepackage[a4paper,top=1cm,bottom=1cm,left=1.2cm,right=1.2cm]{geometry}

\usepackage{amsmath}
\usepackage{lineno}
\usepackage{mathtools}
\usepackage[section]{placeins}
\usepackage{textcomp}
\usepackage{authblk}
\usepackage{xcolor}

\begin{document}

\title{Modal division multiplexing of quantum and classical signals in few-mode fibers}

\author[1]{D. Zia}
\author[2*]{M. Zitelli}
\author[1]{G. Carvacho}
\author[1]{N. Spagnolo}
\author[1]{F. Sciarrino}
\author[2]{S. Wabnitz}
\affil[1]{Physics Department, Sapienza University of Rome, Piazzale Aldo Moro 5, I-00185 Roma, Italy}
\affil[2]{Department of Information Engineering, Electronics, and Telecommunications, Sapienza University of Rome, Via Eudossiana 18, 00184 Rome, Italy}
\affil[*]{mario.zitelli@uniroma1.it}
%\affil[1]{Physics Department, Sapienza University of Rome, Piazzale Aldo Moro 5, I-00185 Roma, Italy}
%\affil[2]{Department of Information Engineering, Electronics, and Telecommunications, Sapienza University of Rome, Via Eudossiana 18, 00184 Rome, Italy}

%\vspace{10pt}

%\begin{document}
\maketitle

 \begin{abstract}
 Mode-division multiplexing using multimode optical fibers has been intensively studied in recent years, in order to alleviate the transmission capacity crunch. %The other big issue related to 
 Moreover, the need for secure information transmission based on quantum encryption protocols %the transmission security 
 leads to investigating the possibility of multiplexing both quantum and classical signals in the same fiber. %channels.
 In this work, we experimentally study the modal multiplexing of both quantum and classical signals at telecom wavelengths, by using a few-mode fiber of 8 km %long of few-mode fiber 
 and modal multiplexers/demultiplexers. We observe %demonstrate 
 the existence of random-mode coupling at the quantum level, leading to cross-talk among both degenerate and non-degenerate channels. Our results demonstrate the feasibility of using few-mode fibers for simultaneously transmitting classical and quantum information, leading to an efficient implementation of physical information encryption protocols in spatial-division multiplexed systems.

 \end{abstract}

%\maketitle

\section{Introduction}

Recent progress in quantum computing \cite{arute2019quantum} has the potential to severely endanger the security of data transmissions. The vast majority of encryption techniques used all over the world are based on mathematical problems which are thought to be computationally hard to solve with a classical computer. However, a quantum computer could easily break such codes \cite{shor1999polynomial,zhong2020quantum}, hence putting at risk most of the current cryptographic infrastructures. In this context, Quantum Key Distribution (QKD) can be an effective solution to the problem. QKD can be used to distribute a key between two parties in such a way that no amount of computational power, classical or quantum, can retrieve any information about it \cite{doi:10.1137/S0097539795293172}. This is possible due to  %can happen thanks
the inherent randomness %and unpredictability that 
provided by quantum physics.  %offers.
In recent years, QKD has seen a significant boost in terms of resources and research invested in the field, achieving secure quantum communication over 600 km \cite{pittaluga2021600}, with key rates %of more 
higher than 10 Mbps, %and 
as well as the implementation of satellite links \cite{liao2017long, PhysRevLett.120.030501}. Another research route that is currently being explored is the miniaturization of QKD terminals%systems
, employing integrated photonic to create compact, scalable, and cost-effective QKD transmitters and receivers \cite{sibson2017chip,paraiso2019modulator,Roger:19}.\\

Most of the research in the field has been focused on the transmission of keys over single-mode fibers (SMF). This is especially true for the cases where long distances are required \cite{pittaluga2021600,PhysRevLett.126.250502,10.1063/1.5027030}, as light traveling through SMF experiences less disruption than in multimode fibers (MMF), due to their large modal dispersion. Nevertheless, most of the local area networks (LANs) deployed in the field, in particular in cloud data centers, employ MMF instead, as they allow for %much 
lower costs than their SMF equivalents in terms of hardware. %This is because their larger core diameter makes it easier to couple light in, allowing cost-effective light sources such as LEDs to be used instead of lasers. 
Both multicore fibers (MCF) \cite{Dynes:16,bacco2019boosting,9102386,xavier2020quantum} and MMF \cite{Namekata:05,Amitonova:20,Wang:20,Zhong:21}, have been tested for %use in 
QKD; however, further research is required to maximize the impact of such an application. Much of the current research in the field has been limited to exploiting the “extremely underfilled mode dispersion” (i.e. by input coupling the QKD signal to the fundamental mode only) of MMF, which makes them behave similarly to SMF \cite{Namekata:05}. Whereas there is a considerable ongoing research and %applicative 
interest in fully exploiting the capacity of both MCF \cite{bacco2019boosting,9102386} and MMF \cite{Wang:20,Zhong:21} by spatial-division-multiplexing (SDM) both classical and quantum signals in different modes of the same fiber. Specifically, in \cite{Wang:20} all-fiber mode-selective couplers (MSC), fabricated by heating and tapering the input/output single mode fibers and the transmission few-mode fiber, were used. However, this method allowed for multiplexing only two nondegenerate fiber modes. 

In this work we make a step forward towards the possibility of massively multiplexing classical and quantum signals over MMF. This is likely to be the most common use case in already-installed networks, where it is not practical to have a dedicated fiber for transmitting QKD signals only. For this aim, here we experimentally demonstrate the simultaneous transmission of quantum and classical signals on a few-mode fiber (FMF).
This is achieved by exploiting the recently developed multiplane light conversion (MPLC) technique \cite{Bade:18}. 
%MPLC permits converting a signal carried by the fundamental mode of a SM fiber into a signal carried by a given mode of MMF. 
The MPLC technique for multiple beam shaping was initially introduced in optical telecommunication applications as a transverse mode multiplexer \cite{fontaine2019,song2021simultaneous,fang2022optical, kupianskyi2023high, goel2023simultaneously}. A MPLC-based spatial multiplexer converts each input multiple single-mode Gaussian beam into: i) the different propagation modes of MMF, such as Laguerre-Gauss (LG), Hermite-Gauss (HG) or ii) any kind of modal base in the case of free-space communications. The MPLC technique has demonstrated its performances for up to 45 HG or LG modes in a standard 50 µm core graded-index MMF. The adaptivity of MPLC to any set of modes of any fiber enabled several fiber capacity increase experiments, where the quality of the spatial modes is critical \cite{Bade:18}. The use of MPLC technology led to achieve the record fiber capacity of 10.16 peta-bit/s, with an aggregate spectral efficiency of 1100 bit/s/Hz \cite{Soma:18}.    

The same MPLC technology also allows for the inverse operation, so that full multiplexing (MUX) and demultiplexing (DeMUX) of many different spatial channels is possible, with relatively low overall insertion losses, which is crucial for quantum optics applications. 
%A further key advantage of the MPLC-based mode MUX-DeMUX technology it its scalability to a large number of modes (up to $>50$ modes have been demonstrated). 
Using a MUX-DeMUX pair in combination with a FMF leads to an effective single-mode path for sub-single-photon quantum pulses to travel through. 
%Hence our approach will pave the way towards a more capillary distribution and deployment of QKD systems in already installed LANs and cloud data center networks, for a cost-effective transition towards a new generation of quantum-safe networks 

As a matter of fact, single photons are an essential resource in quantum information, being used for milestones experiments \cite{zhong2020quantum,aspect1981experimental, boschi1998experimental, bouwmeester1997experimental}. %, and communication \cite{}. 
Indeed, due to their low interaction with the surrounding environment, single photons are a nearly decoherence-free systems, and represent the ideal carrier of information in quantum channels \cite{flamini2018photonic}. Single photon quantum communications allow for safer data transmissions, exploiting free-space links and either SMF or MMF to carry information between the nodes of a network \cite{pittaluga2021600, liao2017long, PhysRevLett.120.030501, cozzolino2019high, nauerth2013air, basso2021quantum, zhang2022future}. Therefore, %experiencing the same issue for the amount of exchanged data of the classical case, 
our MUX-DeMUX approach represents a valuable tool for enhancing the bandwidth of quantum communications.

In this work we demonstrate the feasibility of simultaneously transmitting quantum (i.e., single photon) and classical %(i.e., many photons)  
low-power signals within a MUX-DeMUX system, a configuration which could be implemented in future LANs and cloud data center networks. In particular, this approach may lead to reduced cost and new data security protocols.
Specifically, we experimentally demonstrate the possibility of multiplexing three quantum signals %channels
and two low-power %(i.e., few photons) 
classical signals, %channels,
both at telecom wavelength, into different modal groups of a FMF. We characterized the transmission of single photons over 8-km, %quantum SDM transmission is demonstrated, 
when simultaneously injecting classical signals in the same MUX-DeMUX system. %classical channel transmission. 
The degree of isolation between quantum and classical channels permitted the transmission of signals up to powers compatible with baud rates up to 7.8 Gbaud/s, when several quantum channels are simultaneously transmitted.
In addition, the presence of random-mode coupling (RMC) \cite{Gloge:6774107,Marcuse1973LossesAI,Ho:14} at the quantum level was also studied, and compared with the case of RMC between linear classical channels.

%In practice, the natural candidate to perform the SDM is represented by the Orbital Angular Momentum (OAM) degree of freedom of single photons. The latter is a quantized degree of freedom related to the spatial structure of the field, being associated to the helical shape of beams having a phase term $e^{im\phi}$ where $\phi$ is the azimuthal angle in cylindrical coordinates and $m$ the OAM eigenvalue \cite{allen_0AM_1992, allen1994azimuthal}. 

%In particular, the OAM states can be described using in the orthogonal basis of Laguerre-Gaussian (LG) modes, that therefore are suitable for being used as an information encoding in the SDM. The latter have been used inside quantum protocols to...

%Here, we propose...

%Articoli interessanti demultiplexing: \cite{brandt2020high,fontaine2019laguerre,ji2019high, goel2023simultaneously}.

\section{Experimental Setup}
\begin{figure*}
    \centering
    \includegraphics[width=0.8\textwidth]{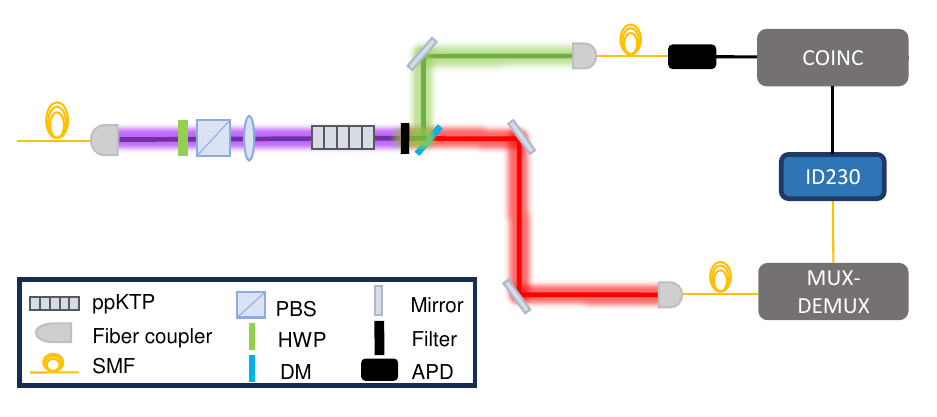}
    \caption{\textbf{Experimental setup.} The CW pump laser at $396$ nm enters the setup through a half-wave plate (HWP) and a polarizing beam splitter (PBS), which are used to control the beam power and set the pump polarization. Next, the laser beam is focused by a lens ($f=25$ cm) on a ppKTP nonlinear crystal. The crystal generates through SPDC a signal photon at $\lambda_s = 532$ nm (green path) and an idler photon at $\lambda_i = 1550$ nm (red path). After splitting signal and idler with a dichroic mirror, the signal is directly revealed by an avalanche photodiode detector (APD), and used as a trigger for the coincidences counter. %\textcolor{red}{While} 
    The IR idler is instead injected into the MUX-FMF-DeMUX system and detected by an infrared single-photon detector (ID230).}
    \label{fig:exp_setup}
\end{figure*}

In our transmission system, the quantum signal is provided by %channels 
 %are formed starting from 
single photon pairs, generated via spontaneous parametric down-conversion (SPDC) through a nonlinear crystal. This is a $6$ mm long periodically-poled Potassium Titanyl Phosphate (ppKTP) crystal, with a poling engineered to produce high energy non-degenerate single photons at two different wavelengths. In particular, when entering the crystal with a continuous-wave (CW) pump laser at a wavelength of $\lambda_p = 396$ nm, the crystal generates an idler photon at $\lambda_i = 1540$ nm and a signal photon at $\lambda_s = 532$ nm, in accordance with energy and momentum conservation (phase-matching) conditions. Our setup is illustrated in Fig. \ref{fig:exp_setup}.
Here, the laser passes through a half-wave plate (HWP) and a polarizing beam splitter (PBS), in order to regulate the power reaching the crystal and entering it with a horizontally polarized pump beam. After the ppKTP, a filter highly attenuates the pump and a dichroic mirror (DM) is used to split the generated photons according to their wavelength, in order to couple them into different SMF. Then the idler is sent to the SDM system (MUX-DeMUX) before entering infrared single-photon detector (ID230).
The signal photon at 532 nm is coupled into a SMF and used as a trigger when reaching the single-photon avalanche photodiode (APD) detector. 

\begin{figure*}
    \centering
    \includegraphics[width=0.6\textwidth]{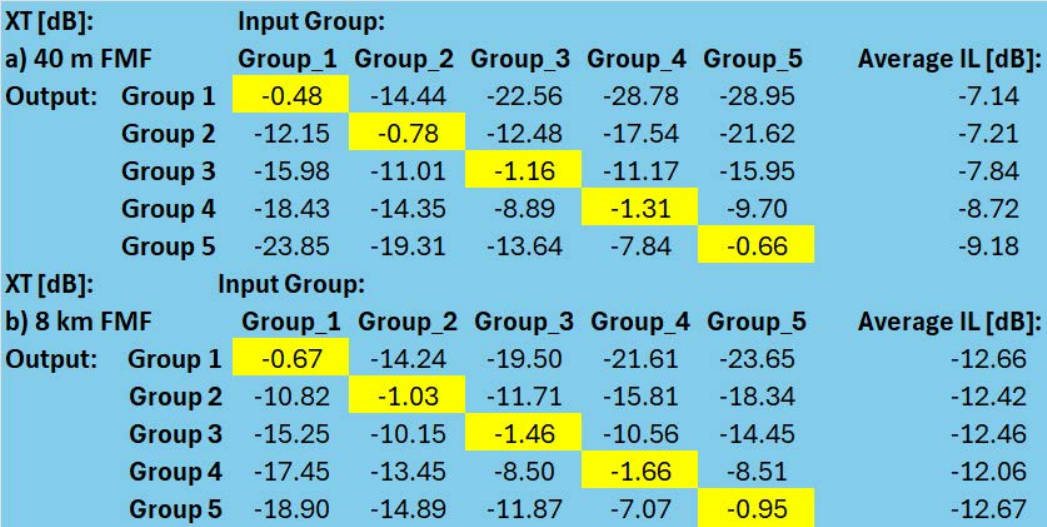}
    \caption{\textbf{Modal group cross-talk [dB]}. In the table are reported the values of the modal cross-talk (in dB) for the MUX-DeMUX system, when a 40 m (a) or 8 km (b) long FMF are used. The results are obtained by adding the powers of the degenerate modes at input and output.}
    \label{fig:XT_table}
\end{figure*}

The SDM system under test is composed by a custom-designed modal MUX-DeMUX (Cailabs PROTEUS-C-15-CUSTOM), which permits to inject and extract the 15 Hermite-Gauss modes $HG_{mn}$ (or $HG_p$, $p=1,..,15$ using a single index) of a 8 km long specialty low-differential-mode-group-delay FMF (Prysmian GI-LP9) \cite{Sillard:16}. The $M=15$ modes per polarization of this fiber consist of $Q=5$ groups of quasi-degenerate modes. We characterized the linear transmission properties of the MUX-DeMUX system using 40 m or 8 km of FMF, coupling 70-fs low power pulses at 1550 nm to the MUX channels, and measuring the output power from the individual outputs of the DeMUX; the channel insertion loss $IL_p$ was measured as the total power at the system output, divided by the input power coupled to a single mode $p$. The linear system performance when using 40 m (8 km) of FMF is the following: the average modal insertion loss at 1550 nm is 8.4 dB (12.4 dB); the worst modal insertion loss is 10.6 dB (14.2 dB); the average cross-talk between different modal groups at 1550 nm is -19.9 dB (-18.1 dB); the worst cross-talk is -12.6 dB (-11.4 dB). By adding the powers of the quasi-degenerate modes at the input and at the output, we obtained the group-by-group cross-talk which is shown in Fig. \ref{fig:XT_table}, after either (a) 40 m or (b) 8 km of FMF, respectively. 

%the linear system performance when using 40 m (8 km) of FMF is the following: average modal insertion loss at 1550 nm: 8.5 (10.3) dB; worst modal insertion loss 9.9 (11.8) dB; average cross-talk between different modal groups at 1550 nm: -15.3 (-15.3) dB; worst cross-talk -10.5 (-10.7) dB.
%\textcolor{red}{potrebbe essere utile fare una tabella con questi valori?}.

\begin{figure*}
    \centering
    \includegraphics[width=0.8\textwidth]{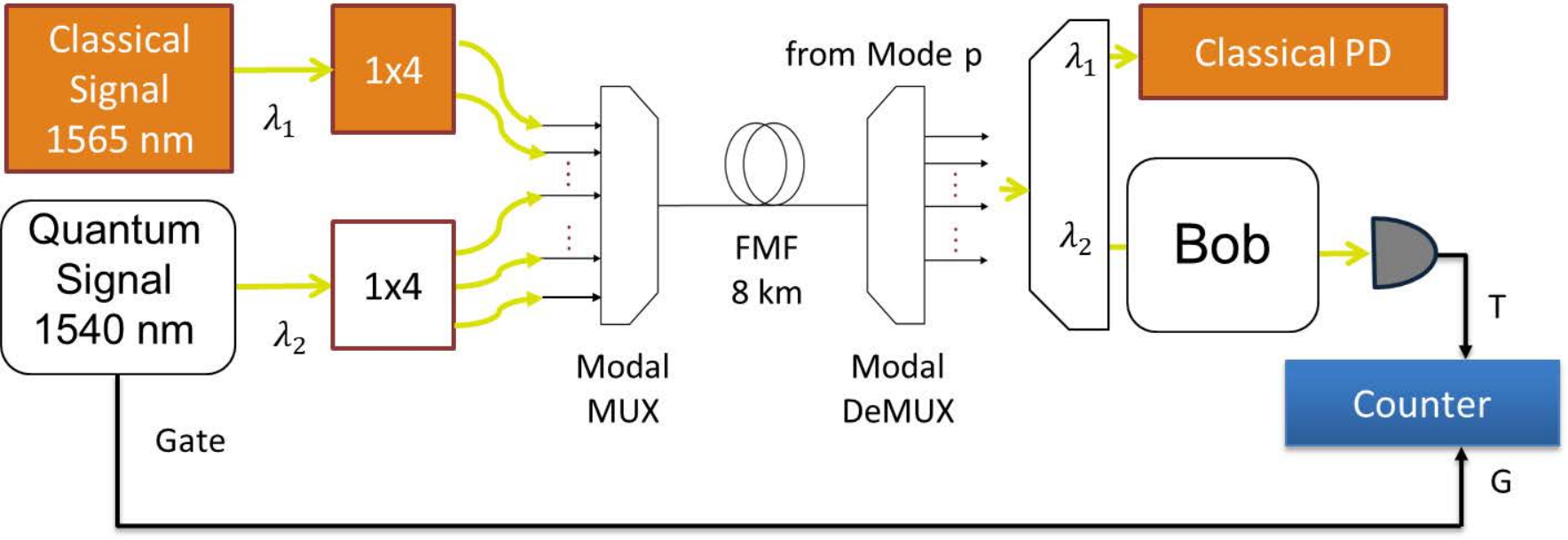}
    \caption{\textbf{Quantum-classical SDM system.} Two classical signals at $\lambda_1=1565$ nm and three quantum signals at $\lambda_2= 1540$ nm are coupled to different modes of a 8 km FMF using an MPLC-based modal multiplexer. At the receiver, channels are space and wavelength-division de-multiplexed. Classical signals are detected by a photodiode PD; quantum channels undergo photon and coincidence counting.
    \label{fig:exp_setup2}}
\end{figure*}

The two classical signals are generated by splitting the output of a CW tunable distributed feedback (DFB) laser emitting at $\lambda_1$=1565 nm, and coupling it to two different FMF modes. Three  %they are wavelength-division multiplexed (WDM) to the 3 
quantum signals at $\lambda_2$=1540 nm are coupled to three FMF modes; these modes are distinct from those used for propagating classical signals. In this way, classical and quantum channels are both SDM and wavelength-division multiplexed (WDM). The 3+2 signals are obtained from 1x4 splitters (Fig. \ref{fig:exp_setup2}). 
The quantum signals %channels
are multiplexed into modes $HG_{00}$ (modal group 1), $HG_{11}$ (group 3) and  $HG_{22}$ (group 5). The classical signals %channels
are multiplexed into modes $HG_{10}$ (group 2) and  $HG_{21}$ (group 4). 
The output is collected from one of the modal channels $p$, which may include both classical and quantum signals at $\lambda_1$ and $\lambda_2$ as a consequence of modal cross-talk; the two wavelengths are de-multiplexed by a wavelength DeMUX provided by a 1-nm bandwidth filter tuned at either 1540 nm or 1565 nm. 

\section{Experimental Results}

At first we studied the transmission properties of the setup by directly injecting single photon states into the MUX-DeMUX system. In particular, using the signal photon as a trigger we observed the coincidence counts between the detectors for several MUX-DeMUX input and output channels. In this way, we could characterize the mode coupling (or cross-talk) between the 15 inputs of the stage, as well as the losses of the setup. 

Classical signals propagating in MMF in the linear regime are affected by modal cross-talk caused by the RMC, which is induced by fiber imperfections, such as micro and macro-bending. Degenerate modes within a modal group are strongly coupled; after few meters, a full power re-distribution between degenerate modes is typically observed. On the other hand, the coupling of modes from different modal groups is much weaker, and is characterized by a coupling coefficient $D$ ranging between $10^{-7}$ and  $10^{-2}$ m$^{-1}$, depending on the fiber type and its bending. RMC leads to a re-distribution of power among non-degenerate modes over distances that may vary between tens of meters up to hundreds of kilometers.
In the linear regime, RMC effects can be modelled by power-flow equations \cite{Savovi2019PowerFI,Gloge:6774107,Olshansky:75}, which predict a diffusion of energy from intermediate groups into both lower-order and high-order modes. These models lead, after a sufficient propagation length, to a steady-state mode power distribution which shows a characteristic decrease of power as the mode order grows larger.
To the best of our knowledge, the effect of RMC at quantum power levels have not yet been investigated.

In our experiments, we measured the coincidence rate $R_{cp}$ between signal and idler photons at the output of a given channel $p$ of the system under test. We acquired data for a time interval of $\Delta t=3$ s, repeated the measurements $N_{test}=100$ times, and obtained the coincidence rates by averaging. The coincidence rate of signal and idler photons at the input of the MUX-DeMUX setup was $R_{in}=2600$ Hz; the tolerance for coincidence detection was $t_c=4.0$ ns. Correspondingly, by using the rate of single photons detected by the APD ($R_{1p}$) at the input and by the ID230 ($R_{2p}$) at the output, the rate of accidental coincidences $R_{ap}=2R_{1p}R_{2p}t_c$ was calculated for each output; next, $R_{ap}$ was subtracted from $R_{cp}$. 

\begin{figure}[h]
\includegraphics[width=0.6\textwidth]{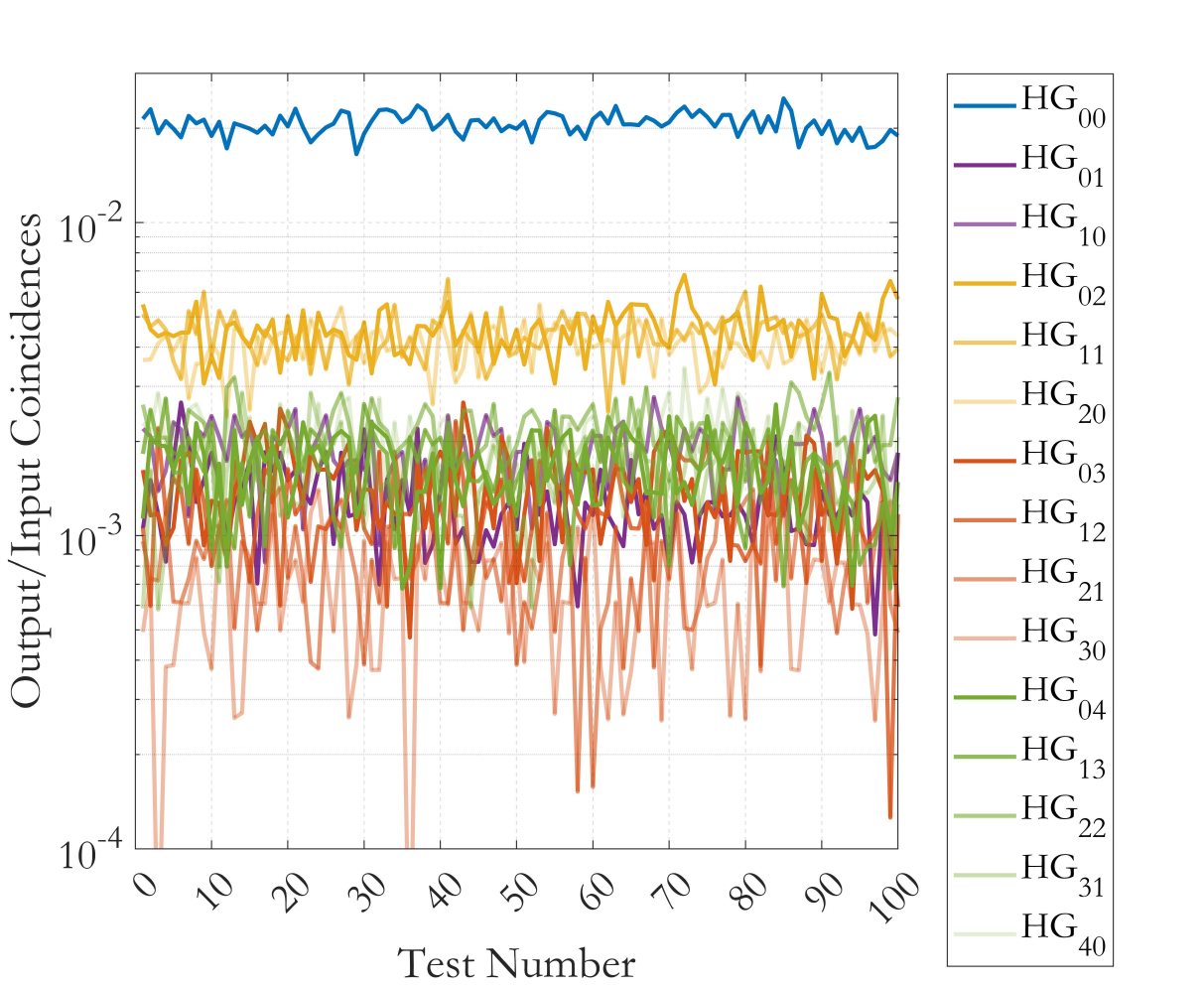}	\centering	
\caption{\textbf{Characterization of system losses.} Measured output-to-input ratio $L_p$ from different modal channels, when the three quantum signals are coupled to modes $HG_{00}, HG_{11}$, $HG_{22}$, respectively with equal average power. Modes of a same group are represented with similar colors.}
%\textcolor{red}{Poco chiaro, forse si può mettere in scala logaritmica per separate ulteriormente i modi? Oppure si può fare un inset che ingrandisce i modi sotto, messi così sono indistinguibili Inoltre magari i segnali dei modi in cui inseriamo i fotoni possiamo metterli come delle spezzate, così che siano distinguibili dagli altri? L'asse verticale potrebbe essere l'efficienza di trasmissione del canale?}}
\label{fig:Fig_CoincidenceFract}
\end{figure}

In Fig. \ref{fig:Fig_CoincidenceFract} we show the measured output-to-input ratio $L_p=(R_{cp}-R_{ap})/R_{in}$ for the different modal channels, when coupling into the MUX-DeMUX the three quantum signals, with no classical signal applied. The error on $L_p$ was calculated from the accidental rate as $\epsilon_{Lp}=R_{ap}/R_{in}$; hence, it is a statistical error normalized in the same way as $L_p$.

\begin{figure}[h]
\includegraphics[width=0.48\textwidth]{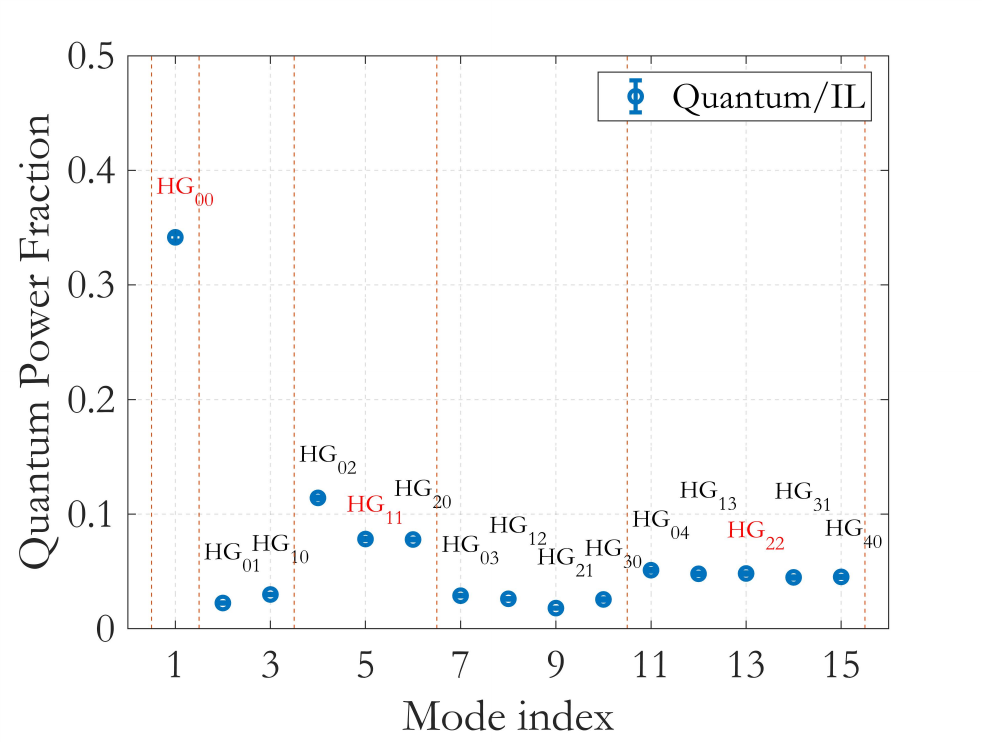}	\centering	
\caption{\textbf{Fractional quantum power.} The $FQP_p$ at the output of the DeMUX, when three equal-power input quantum signals are coupled to modes $HG_{00}$ (group1), $HG_{11}$ (group 3) and $HG_{22}$ (group 5), respectively.}
\label{fig:FigFQP}
\end{figure}

The modal power measured at the output $p$ is affected by both RMC accumulated during transmission, insertion loss and linear cross-talk of the MUX-DeMUX system. In order to study the power re-distribution induced by RMC and other propagation effects, the output power must be normalized to the channel insertion loss $IL_{p}$ (dB), measured in a back-to-back configuration with only 40 m of fiber inserted, by using classical signals at 1550 nm, and supposing that only the insertion loss and cross-talk of the MUX-DeMUX system are present after few meters of fiber. Hence, the fractional quantum power at the output was calculated as

\begin{equation}
FQP_p=\frac{R_{cp}-R_{ap}}{R_{in}10^{IL_p/10}P_{tot}}  ,
\label{eq:FQP}
\end{equation}
with $P_{tot}=\sum_p FQP_p$; the error on the FQP is $\epsilon_{FQP}=\epsilon_{Lp}/(10^{IL_p/10}P_{tot})$. The calculated values for the FQPp for our experimental conditions are illustrated in Fig. \ref{fig:FigFQP}

%Figure \ref{fig:FigFQP} reports the $FQP_p$ with its error at the output of the DeMUX, when 3 equal-power input quantum signals are coupled to the modes $HG_{11}$, $HG_{22}$ and $HG_{22}$ of modal groups 1, 3 and 5, respectively; no classical signals are applied. 
Here, different modal groups are delimited by vertical dashed red lines. Modes excited at the FMF input by quantum channels are highlighted. From the figure, we may observe that, also at the quantum level, RMC re-distributes photons among degenerate modes. Indeed, if no RMC was present, the fractional power measured at the output is expected to be equal to 0.33 only for the excited modes (indicated in red color). To the contrary, our measurements show that power is distributed almost uniformly among all modes of groups 3 and 5 at the system output.

\begin{figure}[h]
\includegraphics[width=0.48\textwidth]{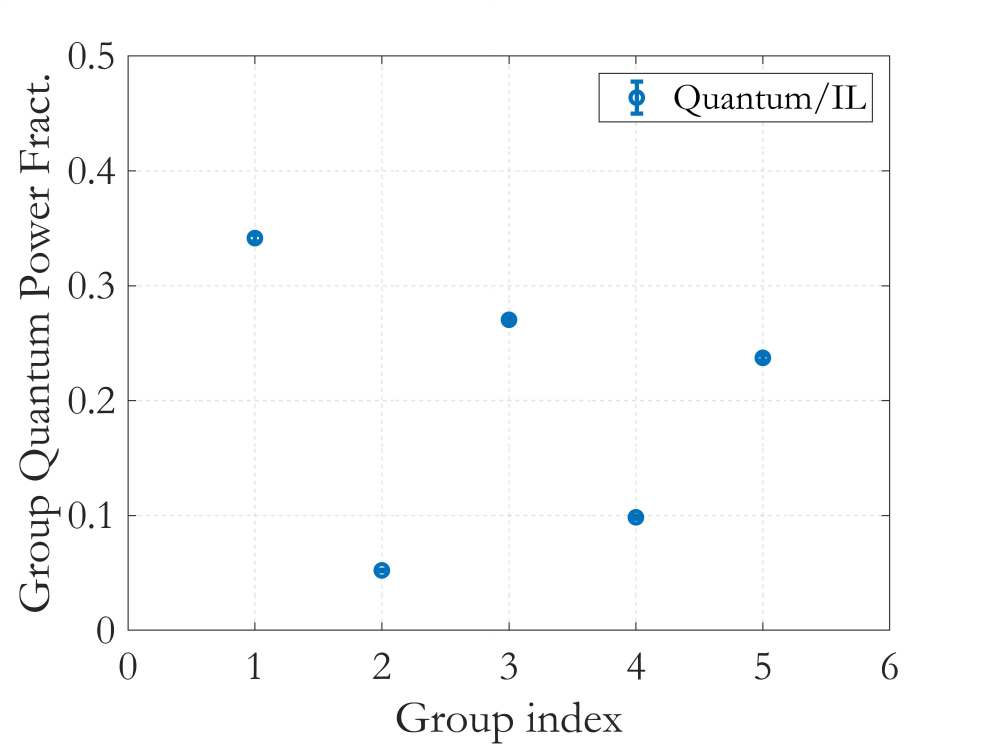}	\centering	
\caption{Measured fractional quantum power of the groups, when the three quantum signals are coupled to modes $HG_{00}, HG_{11}$, $HG_{22}$, respectively, with equal quantum power.}
\label{fig:FigFQPgroups}
\end{figure}

After summing the $FQP_p$ of the degenerate modes, we obtain the fractional quantum power of the groups presented in Fig. \ref{fig:FigFQPgroups}. RMC transfers a 5.2\% fraction of the total power to the second group, which is not carrying a quantum signal, and a 9.8 \% to group 4. The remaining 85 \% share of photons is confined to groups 1, 3 and 5 which carry quantum signals, with power fractions ranging between 24 \% and 34 \%. Figure \ref{fig:FigFQPgroups} shows that RMC tends to cumulate power to the lower-order modes.
% \textcolor{red}{mettiamo valore medio?}.

In a second test, two classical CW signals at $\lambda_2=1565$ nm were coupled to modes $HG_{10}$ (group 2) and $HG_{21}$ (group 4) with variable power $P$ per channel, simultaneously with quantum signals coupled to modes $HG_{00}$ (group 1) and $HG_{11}$ (group 3), respectively, at 1540 nm. 
RMC cumulated during transmission and linear cross-talk of the MUX-DeMUX system (-18.1 dB on average) mixes photons of classical and quantum signals; hence, it was necessary to wavelength-division-multiplex the SDM classical and quantum channels, so that they could be separated at the system output by a bandpass filter, as it is done when simultaneously transmitting classical and quantum channels in SMF systems.
%Because of the substantial linear cross-talk between different mode groups at the system output (-15.3 dB on average), it was necessary to wavelength-division-multiplex the SDM classical and quantum channels, so that they could be separated at the system output by a bandpass filter, as it is done when simultaneously transmitting classical and quantum channels in SMF systems. 

Still, a cross-talk was observed at the output of a channel carrying the quantum signal, in the form of an increase of the total number of coincidences owing to accidental coincidences between quantum and classical signals. 
The classical and quantum signals interfere because of the limited extinction ratio of the modal DeMux and the wavelength-division demultiplexing filter.
For increasing power of the input classical signals, it was measured the rate of coincidences between quantum signal at the input and output modes.
%The rates of coincidence were measured for increasing power of the input classical signals at the output of the channels carrying the single photons signal. 
Coincidence rates were normalized to the insertion loss $IL_p$.
The signal-to-noise ratio of quantum to classical signals was calculated as follows:

\begin{equation}
SNR_p=10 \log_{10} \Big[ \frac{R_{cp}(0)-R_{ap}}{R_{cp}(P)-R_{cp}(0)} \Big] \simeq 10 \log_{10} \Big[ \frac{R_{cp}(0)}{R_{cp}(P)-R_{cp}(0)} \Big],
\label{eq:SNR}
\end{equation}

with $R_{cp}(P)$ the rate of coincidences from output mode $p$ when classical signals with power $P$ are applied to groups 2 and 4.

\begin{figure}[h]
\includegraphics[width=0.48\textwidth]{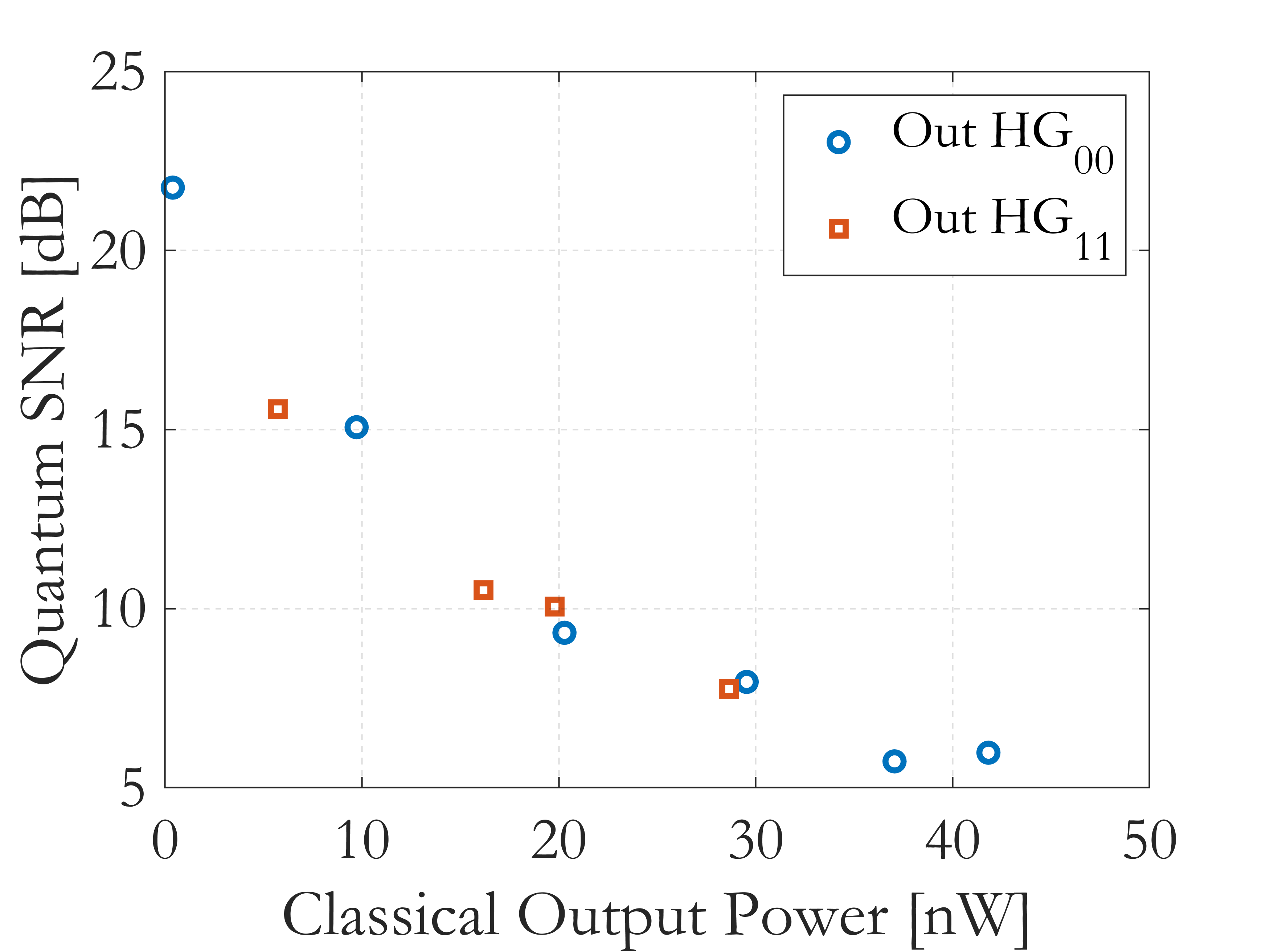}	\centering	
\caption{\textbf{Quantum signal-to-noise ratio.} Measured signal-to-noise ratios for quantum signals $HG_{00}$ and $HG_{11}$ vs. classical output power $P$.}
\label{fig:SNR}
\end{figure}

Figure \ref{fig:SNR} shows the measured $SNR_p$ from output modes $HG_{00}$ and $HG_{11}$, for increasing values of the power $P$ of classical channels. After 8 km transmission, $SNR_p$ remains larger than 10 dB for classical signal output powers up to $P_c$=20 nW.
Assuming Poisson statistics for the photons in one bit pulse of a classical signal, the quantum limit (i.e., for an ideal shot-noise limited detector, neglecting the presence of thermal noise) to obtain a bit-error-rate $BER < 10^{-9}$ is $N_p$=20 photons per pulse \cite{agrawal2012fiber}.
The corresponding maximum baud rate $B$, or pulse repetition frequency for a classical signal would be

\begin{equation}
B \leq \frac{P_c}{N_p h \nu}.
\label{eq:Baud}
\end{equation}

In our case, this corresponds to a maximum baud rate of $B \leq 7.8$ Gbaud/s per classical channel, in the presence of quantum signals.

\FloatBarrier

\section{Conclusions}

In this work, we have characterized the impact of linear RMC among modes of a FMF when quantum signals are propagated. Photons are exchanged rapidly among quasi-degenerate modes, resulting into their statistical equipartition among the modes of a given group. Weaker power transfer occurs between different modal groups, as predicted by the power-flow theory \cite{Gloge:6774107,Savovi2019PowerFI}. A cross-talk of less than 10 \% (-10 dB) was observed among quantum signals after 8 km transmission, which may permit the modal multiplexing of quantum signals.

When classical channels were added via both modal and wavelength division multiplexing, an additional cross-talk was observed in terms of accidental quantum-classical coincidences. The quantum $SNR_p$ remains higher than 10 dB for classical channel output powers up to 20 nW, after transmission over 8 km of FMF. This could correspond to a limit of 7.8 Gbaud/s to the pulse repetition rate of the classical signals in the presence of shot noise, when multiplexed to quantum signals. 
%By using a multilevel modulation of real classical signals, which permits a capacity up to 10 bit/pulse, the classical rate could arrive to 78 Gbit/s when quantum signals are present.

\section{Acknowledgments/Funding}
We thank Cailabs for providing us with the MUX/DeMUX system used in our experiments. Project ECS 0000024 Rome Technopole, Funded by the European Union – NextGenerationEU, and Horizon Europe ERC PoC (no. 101081871).

%\bibliographystyle{unsrt}
%\bibliography{biblio.bib}

\end{document}